\documentclass{article}
\usepackage{smc2020}
\usepackage{times}
\usepackage{ifpdf}
\usepackage[english]{babel}
\usepackage{cite}
\usepackage{multirow, booktabs, graphicx, subcaption}
\usepackage{soul, stfloats}

\def\papertitle{Evaluation of CNN-based Automatic Music Tagging Models}
\def\firstauthor{Minz Won}
\def\secondauthor{Andres Ferraro}
\def\thirdauthor{Dmitry Bogdanov}
\def\fourthauthor{Xavier Serra}

\newif\ifpdf
\ifx\pdfoutput\relax
\else
  \ifcase\pdfoutput
      \pdffalse
  \else
      \pdftrue
\fi

\ifpdf 
  \usepackage[pdftex,
    pdftitle={\papertitle},
    pdfauthor={\firstauthor, \secondauthor, \thirdauthor, \fourthauthor},
    bookmarksnumbered, 
    pdfstartview=XYZ 
  ]{hyperref}

  \usepackage[figure,table]{hypcap}

\else 
  \usepackage[dvips,
    bookmarksnumbered, 
    pdfstartview=XYZ 
  ]{hyperref}  

  \usepackage[dvips]{epsfig,graphicx}
  \graphicspath{{./figures/}}
  \DeclareGraphicsExtensions{.eps}

  \usepackage[figure,table]{hypcap}
\fi

\hypersetup{
    colorlinks,%
    citecolor=black,%
    filecolor=black,%
    linkcolor=black,%
    urlcolor=black
}

\title{\papertitle}

 \fourauthors
  {\firstauthor} 
  {Music Technology Group \\
  Universitat Pompeu Fabra, Barcelona\\
    {\tt \href{mailto:minz.won@upf.edu}{\{firstname.lastname\}@upf.edu}}}
  {\secondauthor} 
  {\thirdauthor} 
  {\fourthauthor} 

\begin{document}

\capstartfalse
\maketitle
\capstarttrue

\begin{abstract}
Recent advances in deep learning accelerated the development of content-based automatic music tagging systems. Music information retrieval (MIR) researchers proposed various architecture designs, mainly based on convolutional neural networks (CNNs), that achieve state-of-the-art results in this multi-label binary classification task. However, due to the differences in experimental setups followed by researchers, such as using different dataset splits and software versions for evaluation, it is difficult to compare the proposed architectures directly with each other. To facilitate further research, in this paper we conduct a consistent evaluation of different music tagging models on three datasets (MagnaTagATune, Million Song Dataset, and MTG-Jamendo) and provide reference results using common evaluation metrics (ROC-AUC and PR-AUC). Furthermore, all the models are evaluated with perturbed inputs to investigate the generalization capabilities concerning time stretch, pitch shift, dynamic range compression, and addition of white noise. For reproducibility, we provide the PyTorch implementations with the pre-trained models.

\end{abstract}
%


\section{Introduction}\label{sec:introduction}
Automatic music tagging is a multi-label binary classification task that aims to predict relevant tags for a given song. 
Typically these tags carry useful semantic music information that can later be used for other applications such as music recommendation or music retrieval. 
To tackle the problem of music tagging, recent studies in music information retrieval (MIR) adopted deep neural networks, mostly based on convolutional neural networks (CNNs), motivated by their huge success in other domains (e.g., computer vision, natural language processing). The introduction of deep learning helped to break the previous glass ceiling in the performance of music tagging systems and 
MIR researchers started actively proposing ingenious architecture designs. As a result, the hand-crafted feature-based approaches were replaced by data-driven feature learning approaches in most recent automatic music tagging research. 

To maximize the advantages of CNNs, a deep fully convolutional network (FCN) was proposed for music tagging \cite{choi2016automatic}. It uses a stack of $3 \times 3$ rectangular filters followed by a max-pooling layer.
As an alternative, the Musicnn \cite{pons2018end}, also a Mel-spectrogram-based CNN, tried to incorporate domain knowledge into its filter designs so that the model can capture timbral characteristics and temporal patterns using vertical filters and horizontal filters, respectively. Sample-level CNN \cite{lee2017sample} pursued an assumption-free end-to-end model by applying 1D convolution directly to a raw audio waveform, and the following research \cite{kim2018sample} improved the performance by adding squeeze-and-excitation blocks \cite{hu2018squeeze}. Different from images, however, music is sequential. For this reason, a convolutional recurrent neural network (CRNN) \cite{choi2017convolutional} was proposed to extract local features using CNNs and summarize them with recurrent neural networks (RNNs). Another sequence modeling approach \cite{won2019toward} adapted the self-attention mechanism \cite{devlin2018bert} to summarize the temporal sequence of the extracted local features by CNNs. Finally, Harmonic CNN \cite{won2020data} used a harmonically stacked trainable representation to preserve spectro-temporal locality in convolution layers.

Unfortunately, due to different experimental setups followed by the authors of these approaches when reporting results (e.g., dataset splits, library versions, computing environments, and optimization methods), it is difficult to compare the proposed architectures directly with each other. In this paper, we address this issue and report experimental results for various state-of-the-art music tagging models using three different datasets (MagnaTagATune, Million Song Dataset, and MTG-Jamendo dataset) with a consistent experimental setup. In addition, we conduct experiments to assess the robustness of these architectures against four different types of deformations~\cite{mcfee2015software} and determine their generalization abilities. For the reproducibility, we provide PyTorch implementations for all the models considered and their pre-trained models.\footnote{https://github.com/minzwon/sota-music-tagging-models/}

The paper is organized as follows. Section 2 describes automatic music tagging tasks with details of multiple instance problem, evaluation metrics, and dataset descriptions. Section 3 elaborates on the model designs. We report model performances and their generalization capabilities in Section 4 and Section 5, respectively. Finally, Section 6 concludes the paper.

\section{Automatic Music Tagging}\label{sec:MusicTagging}
\subsection{Multiple Instance Problem}
Various semantic information characterizing music, such as genres, subgenres, moods, instruments, decades, and languages, can be expressed in the form of music tags.
Automatic content-based music tagging is a task that aims to predict such relevant tags for a given song based on its acoustic characteristics. However, depending on the tag, relevant characteristics are not necessarily predominant in the entire music recording. For example, when a song has a tag \textit{female vocal}, it does not imply that the female vocal appears in every time segment of the song. 
In essence, this is a multiple instance problem~\cite{dietterich1997solving}. A song can have many acoustic characteristics (instances) that describe it, but often only specific characteristics (instances) are responsible for the associated music tag. In most cases, we do not have time precise instance-level annotations because the precise labeling can be laborsome, therefore a music tag associated with a song is simply applied to all music excerpts (instances) of the song during training~\cite{mcfee2018adaptive}.



There are two approaches to handle the multiple instance problem in music tagging. One is to train the model on full songs and produce song-level predictions from a song-level input. 
From a given song-level input, the model has to predict relevant tags. Another one is to train the model on short audio chunks (instances) and generate chunk-level predictions, which can be later aggregated (e.g., majority vote, average) in the evaluation phase.
In this paper, we refer to these two approaches as song-level training and chunk-level training, respectively.

\subsection{Evaluation Metrics}
A common evaluation metric of binary decision problems is the area under receiver operating characteristic curve (ROC-AUC). However, the area under precision-recall curve (PR-AUC) can be more informative for evaluation on highly skewed datasets \cite{davis2006relationship}. Hence, we use both macro ROC-AUC and PR-AUC to evaluate all considered music tagging models with both scores being averaged across all the tags the models operate with. 
For our robustness studies, we report the variance of ROC-AUC and PR-AUC scores obtained on perturbed audio inputs.

\subsection{Datasets}

\noindent\textbf{MagnaTagATune (MTAT)} \cite{law2009evaluation} is one of the most commonly used datasets for benchmarking automatic music tagging systems. It contains multi-label annotations by genre, mood, and instrumentation for 25,877 audio segments, each 30s long. The audio is in the MP3 format (32 Kbps bitrate and 16 kHz sample rate). Originally the dataset is split into 16 folders, and commonly the first 12 folders are used for training, the 13th for validation, and the last three are used for testing. Only 50 most frequent tags are typically used for the task. The top 50 tags include genre and instrumentation labels, as well as decades (e.g., '80s' and '90s') and moods.

In our work, we follow the same data split and we use the top 50 tags to be consistent with results reported in the majority of previous studies.\footnote{\url{https://github.com/jongpillee/music_dataset_split/tree/master/MTAT_split}} However, we have noted that the performances reported in some studies are inconsistent, as their authors discarded audio segments without any associated tags (leading to slightly higher values of performance). 
Also, some of the previous studies are re-using those reports unintentionally compare incompatible performance values, which is one of the main motivations of our work.


\noindent\textbf{Million Song Dataset (MSD)}~\cite{bertin2011million} is a dataset of audio features for one million songs, partially expanded by the MIR community with 
crowdsourced tags 
from \textit{Last.fm} as well as 
a mapping to 30s audio preview segments originally obtained from \textit{7digital}.\footnote{\url{https://www.7digital.com}}  
In total, this subset of the dataset contains 241,904 annotated song segments and it is commonly used for benchmarking on a larger scale. The tags cover genre, instrumentation, moods and decades. The audio segments vary in quality, being encoded as MP3s with a bitrate from 64 to 128 Kbps and sample rate of 22 kHz or 44 kHz.

Again, in our study we follow the dataset split commonly used by researchers\footnote{\url{https://github.com/jongpillee/music_dataset_split}}~\cite{lee2017sample,pons2018end} and use only 50 most frequent tags for consistency with previous studies. This split includes 201{,}680 songs for training, 11{,}774 for validation and 28{,}435 for testing. 
Unfortunately, similarly to MTAT, some previous studies report inconsistent comparisons of automatic tagging approaches, as there also exists a slightly different split,\footnote{\url{https://github.com/keunwoochoi/MSD_split_for_tagging}} containing audio segments of shorter duration and on which lower performance values were reported. 
Note that the tag annotations available for this dataset are inherently noisy as they come from a free-form social tagging application for music enthusiasts and are used without any preprocessing intended to improve the quality of tags \cite{choi2017effects}.


\noindent\textbf{MTG-Jamendo Dataset} \cite{bogdanov2019mtg} contains audio for 55,701 full songs and is built using music publicly available on the \textit{Jamendo}\footnote{\url{https://jamendo.com}} music platform under Creative Commons licenses.
The minimum duration of each song is 30s, and they are provided in the MP3 (320 Kbps bitrate). Thus, this dataset contains significantly larger audio segments with higher encoding quality than MTAT and MSD. The tracks in the dataset are annotated by 692 different tags covering genres, instrumentation, moods and themes. All tags were originally provided by the artists submitting music to Jamendo, but they were preprocessed with the goal of tag cleaning by the creators of the dataset.

Multiple splits of the data are provided for training, validation and test. In this work we use the split-0 
and 50 most frequent tags.\footnote{\url{https://github.com/MTG/mtg-jamendo-dataset}} 
As this dataset has been released only recently, not many studies are reporting the performance of the models using it yet. Yet, it is a useful addition for the evaluation methodologies followed by researchers in order to better assess the generalization of their models. 


\section{Models}\label{sec:models}
\begin{figure*}[t!]
    \centering
    \includegraphics[width=0.9\linewidth]{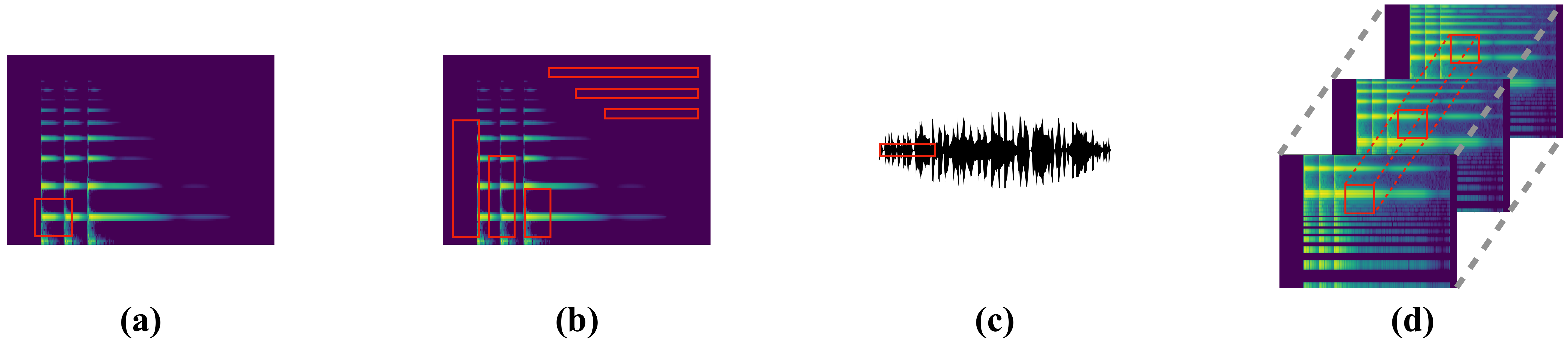} 
    \caption{Shapes of the first convolution filters and input representations of different models. (a) FCN, CRNN, self-attention, and short-chunk CNN with a Mel spectrogram input (b) Musicnn with a Mel spectrogram input (c) sample-level CNN with a raw audio input (d) Harmonic CNN with a stacked harmonic tensor.}
    \label{fig:models}
\end{figure*}

We here describe the architectures of all models considered in this study. \textit{Essentia} \cite{bogdanov2013essentia} and \textit{Librosa} \cite{mcfee2015librosa} libraries are used for loading audio files and extracting Mel spectrograms, respectively. We re-sampled the audio to 16 kHz sample rate. 
Table~\ref{tab:rep} shows different input lengths and the number of Mel bands. 
Since some models have a smaller number of Mel bands in their original implementations, we further experimented with a larger number of Mel bands for the fair comparison (marked with \mbox{*}).
All Mel-spectrogram-based approaches used 512-point FFT with a 50\% overlap.
For training the models, we used a unified optimization method: a mixture of scheduled ADAM \cite{kingma2014adam} and stochastic gradient descent (SGD), introduced in \cite{won2019toward}. The best model is selected based on the validation loss. In the evaluation, chunk-level models average predictions over 16 chunks to perform the final prediction. Note that FCN and CRNN are song-level training models for MTAT and MSD but they perform as chunk-level training models for the MTG-Jamendo dataset since the songs in the dataset are longer than 29.1s which is the size of receptive fields of both models.

\begin{table}[t!]
\centering
\footnotesize
\begin{tabular}{@{}l@{}r@{\hskip 0.1in}r@{\hskip 0.1in}r@{}}
\toprule
Model &input length &\# of Mel &training\\ \midrule
FCN &29.1s  &96 &song-level\\
FCN (128)$^{*}$ &29.1s  &128 &song-level\\
Musicnn &3s &96 &chunk-level\\
Musicnn (128)$^{*}$ &3s &128 &chunk-level\\
Sample-level CNN&3.69s  &-  &chunk-level\\
CRNN    &29.1s  &96 &song-level\\
CRNN (128)$^{*}$    &29.1s  &128 &song-level\\
Self-attention  &15s    &128  &chunk-level\\
Harmonic CNN    &5s &128  &chunk-level\\
Short-chunk CNN&3.69s    &128  &chunk-level\\ \bottomrule
\end{tabular}
\caption{Experimental setups of different models.}
\label{tab:rep}
\end{table}


\subsection{Fully Convolutional Network}\label{sec:FCN}
A fully convolutional network (FCN) is a variant of CNN that consists of only convolutional layers without any fully-connected layers. A FCN for music tagging \cite{choi2016automatic} uses Mel spectrogram inputs. In the preprocessing step, a 29.1s audio segment is converted to a $96 \times 1366$ Mel spectrogram. It is then used as an input and is passed through 4 convolutional layers. Each convolutional layer uses homogeneous $3 \times 3$ 2D filters (Figure \ref{fig:models}-(a)) followed by a max-pooling layer. Different sizes of strides are used for max-pooling layers ((2, 4), (4, 5), (3, 8), (4, 8)) to increase the size of receptive fields to cover the entire input Mel spectrogram ($96 \times 1366$). In the original paper, FCN was trained with a song-level training method since the track durations in MTAT correspond to the size of the receptive field. 
However, this is not the case for MTG-Jamendo containing longer tracks, where chunk-level (29.1s) training is applied.


\subsection{Musicnn}\label{sec:Musicnn}
The Musicnn \cite{pons2018end} model also uses Mel spectrograms as its inputs. 
Different from previously proposed models, the architecture design choices in Musicnn rely on some intuition from the music domain knowledge.
The first convolutional layer of Musicnn consists of vertical and horizontal filters. Vertical filters are designed to capture pitch-invariant timbral features (bottom-left of Figure \ref{fig:models}-(b)): e.g., $38 \times 7$ filter captures sub-band information of short period of time. To enforce the pitch-invariancy, the following max-pooling layer pools the maximum value across the frequency axis. Horizontal filters, on the other hand, capture temporal energy envelope of the audio. 
After the mean-pooling across the frequency axis of input Mel spectrograms, horizontally long filters (e.g., $1 \times 165$) capture the temporal energy patterns (top-right of Figure \ref{fig:models}-(b)). The extracted timbral and temporal features are concatenated in the channel, then the following 1D convolutional layers summarize them to predict relevant tags. Different from FCN, the Musicnn only uses short audio excerpts (3s) as its inputs during training, i.e., chunk-level training.

\subsection{Sample-level CNN}
Sample-level CNN \cite{lee2017sample} tackles the automatic music tagging problem in an end-to-end fashion. It takes raw audio waveforms as its inputs. Sample-level CNN is simpler and deeper than Mel spectrogram-based approaches. It consists of ten 1D convolutional layers with $1 \times 3$ filters and $1 \times 3$ max-poolings (Figure \ref{fig:models}-(c)). Trained front-end filters perform similar to the process of deriving Mel spectrograms and the back-end convolution layers summarize them. 
We also 
considered a variation of sample-level CNN \cite{kim2018sample} with squeeze-and-excitation (SE) \cite{hu2018squeeze} blocks. Sample-level CNN and its variant with SE blocks also use short audio excerpts (3.69s) for the chunk-level training.

\subsection{Convolutional Recurrent Neural Network}
Convolutional recurrent neural network (CRNN) \cite{choi2017convolutional} uses Mel spectrogram inputs. A CRNN can be described as a combination of CNNs and RNNs. The CNN front end extracts local features and the RNN back end summarizes them temporally. Since RNNs are more flexible than CNNs for summarizing sequential information, it can be beneficial to use RNNs for predicting tags that may be affected by global structures (e.g., moods/themes). Four convolutional layers with $3 \times 3$ 2D filters are used in the front end and two-layer RNNs with gated recurrent units (GRU) are used in the back end. Long music excerpts (29.1s) are used as inputs of CRNN.

\subsection{Self-attention}
The self-attention-based music tagging model \cite{won2019toward} shares the same intuition as CRNN to extract local features with CNNs and summarize them with sequence models. The only difference is that the self-attention mechanism is used instead of the RNNs for the temporal summarization back end. Motivated by its huge success in natural language processing \cite{devlin2018bert}, the authors adapted the Transformer encoder, which is a deep stack of self-attention layers, for automatic music tagging. 15s-long audio excerpts are used for training the self-attention model.

\subsection{Harmonic CNN}
Harmonic CNN \cite{won2020data} takes advantage of trainable band-pass filters and harmonically stacked time-frequency representation inputs. Trainable filters (mainly trainable bandwidths) bring more flexibility to the model. And harmonically stacked representation preserves spectro-temporal locality while keeping the harmonic structures through the channel of the input tensor in the first convolution layer (Figure \ref{fig:models}-(d)) as introduced in \cite{bittner2017deep}. The number of trainable frequency bands is set to 128 and the number of harmonics considered for stacking is 6. Chunk-level training with 5s audio segments is performed.

\subsection{Short-chunk CNN}
According to 
the previous 
work~\cite{won2020data}, a simple 2D CNN with $3 \times 3$ filters can already claim exceptional results when it is trained with short chunks of audio, i.e., chunk-level training. It is a  very prevalent type of CNN (sometimes referred as \textit{vgg}-like) but, to the best of our knowledge, 
there are no references for this architecture design in music tagging research. Hence, we implemented a 7-layer CNN with a fully-connected layer and its extension with residual connections \cite{he2016deep}. Different from FCN, it uses a smaller size of max-pooling ($2 \times 2$) because the input segment is way shorter than the song-level inputs (29.1s). We used 128 Mel bins so that 7 max-pooling layers can summarize them into a single dimension ($2^7=128$). It uses 3.69s audio excerpts, hence we call this model ``short-chunk CNN" in this paper to differentiate it from FCN.

\section{Results}\label{sec:results}
\begin{table*}[ht!]
\centering
\small
\begin{tabular}{@{}lcccccc@{}}
\toprule
\multicolumn{1}{c}{\multirow{2}{*}{Methods}} & \multicolumn{2}{c}{MTAT}    & \multicolumn{2}{c}{MSD} & \multicolumn{2}{c}{MTG-Jamendo} \\ \cmidrule(l){2-7} 
\multicolumn{1}{c}{}    & ROC-AUC   & PR-AUC    & ROC-AUC   & PR-AUC    & ROC-AUC   & PR-AUC   \\ \midrule
FCN \cite{choi2016automatic}                    & 0.9005    & 0.4295    & 0.8744    & 0.2970    & 0.8255    & 0.2801    \\
FCN (with 128 Mel bins)                  & 0.8994    & 0.4236    & 0.8742    & 0.2963    & 0.8245    & 0.2792    \\
Musicnn \cite{pons2018end}                      & 0.9106    & 0.4493    & 0.8803    & 0.2983    & 0.8226    & 0.2713    \\
Musicnn (with 128 Mel bins)                      & 0.9092    & 0.4546    & 0.8788    & 3036    & 0.8275    & 0.2810    \\
Sample-level \cite{lee2017sample}               & 0.9058    & 0.4422    & 0.8789    & 0.2959    & 0.8208    & 0.2742    \\
Sample-level + SE \cite{kim2018sample}          & 0.9103    & 0.4520    & 0.8838    & 0.3109    & 0.8233    & 0.2784    \\
CRNN \cite{choi2017convolutional}               & 0.8722    & 0.3625    & 0.8499    & 0.2469    & 0.7978    & 0.2358    \\
CRNN (with 128 Mel bins)              & 0.8703    & 0.3601    & 0.8460    & 0.2330    & 0.7984    & 0.2378    \\
Self-attention \cite{won2019toward}             & 0.9077    & 0.4445    & 0.8810    & 0.3103    & 0.8261    & 0.2883    \\
Harmonic CNN \cite{won2020data}                 & 0.9127    & 0.4611    & \textbf{0.8898}    & \textbf{0.3298}    & 0.8322    & 0.2956         \\
Short-chunk CNN                            & 0.9126    & 0.4590         & 0.8883    & 0.3251    & \textbf{0.8324}    & \textbf{0.2976}         \\
Short-chunk CNN + Res                        & \textbf{0.9129}    & \textbf{0.4614}    & \textbf{0.8898}    & 0.3280    & 0.8316    & 0.2951    
\\ \bottomrule
\end{tabular}
\caption{Performances of state-of-the-art models.}
\label{tab:mainresults}
\end{table*}

We report ROC-AUC and PR-AUC of all implemented models using three datasets in Table \ref{tab:mainresults}. In general, models trained with short audio excerpts (Musicnn, variants of sample-level CNN, Self-attention, Harmonic CNN, variants of short-chunk CNN) outperform other models trained with relatively longer audio segments (FCN, CRNN). Training with short chunks (instances) is noisier: e.g., an audio excerpt can have a tag \textit{guitar} if a guitar appears in the song even though the selected excerpt doesn't include guitar sound in it. However, one can expect a much larger number of examples during the training (e.g., 25,877 tracks $\times$ 16 chunks = 414,032 examples). We suspect this brings the performance gain when the model is trained with short chunk-level examples. Furthermore, most of the top 50 tags in the three datasets can be identified only with a short audio excerpt (e.g., instruments, genres). Thus, the model does not need a long sequence of audio to perform its binary classification task. For the top 50 tags in each dataset we experimented with, it is more beneficial to use chunk-level training with short audio excerpts than the song-level training.

Short-chunk CNN, short-chunk CNN with residual connections, and Harmonic CNN showed the best results for every dataset. These three models are trained 
on short audio excerpts (3.69s or 5s) and they use $3 \times 3$ convolutional filters followed by $2 \times 2$ max-poolings. 
FCN uses similar filters, but larger max-poolings which increase its size of the receptive field to fit long audio segments (29.1s). We conclude that smaller max-poolings with shorter audio excerpts work better for CNNs with $3 \times 3$ filters.

Musicnn shows competitive results in MTAT. However, other models (sample-level + SE, self-attention) outperform Musicnn on larger datasets (MSD and MTG-Jamendo). This confirms an intuition that domain knowledge can be beneficial for relatively small datasets, reported in \cite{pons2018end}. However, the design choices for parameters of Musicnn restricts the power of the model when it is trained on larger datasets.

For the sequential models, self-attention outperforms the CRNN. Different from self-attention mechanisms, RNNs with long sequence inputs suffer from vanishing gradient problems. Self-attention mechanism alleviates the problems by providing direct paths between all time steps. According to the reported visualizations \cite{won2019toward}, self-attention performs well for pinpointing relevant short-time acoustic features in the audio sequence, but it was difficult to determine if the model learned long-time characteristics properly. To determine such abilities, some tags related to a global structure have to be cherry-picked and evaluated.

Since FCN, Musicnn, and CRNN use Mel spectrogram inputs with 96 Mel bands, there can be relative disadvantages when they are compared with other models using 128 Mel bands. For the fair comparison, we experimented with FCN, Musicnn, and CRNN using 128 Mel bands. A larger number of Mel bands did not show any significant impacts on the performances. Since each architecture design was optimized for a smaller number of Mel bands, simply increasing the size of input Mel bands cannot guarantee the optimized performance of the models.

\section{Robustness Studies}\label{sec:robustness}
\subsection{Input Deformations}
To further investigate the performance of different state-of-the-art models, we conducted robustness studies. If a pre-trained model has good generalization abilities, the prediction of the model should not be sensitive against small perturbations in the input audio. By applying four different audio deformations to the test set (pitch shift, time stretch, dynamic range compression, and addition of white noise), we intended to determine the generalization abilities of the models. Note that we applied these four deformations only to the test set, which means that the models have never been exposed to the same deformations during training. All employed deformations are based on an existing music data augmentation framework (MUDA)\footnote{\url{https://github.com/bmcfee/muda}} \cite{mcfee2015software}:

\noindent\textbf{Pitch shift} by $n \in \{-1, 1\}$ semitones.

\noindent\textbf{Time stretch} by $\gamma \in \{2^{-1/2}, 2^{1/2}\}$.

\noindent\textbf{Dynamic range compression} following \textit{speech} and \textit{music (standard)} settings of Dolby E standards \cite{dolby2002standards}.

\noindent\textbf{White noise addition} $x_{mixed} = (1 - \alpha) \cdot x + \alpha \cdot x_{noise}$ where $\alpha \in \{0.1, 0.4\}$.


\subsection{Robustness results}
\begin{figure*}[t!]
    \centering
    \includegraphics[width=1.0\linewidth]{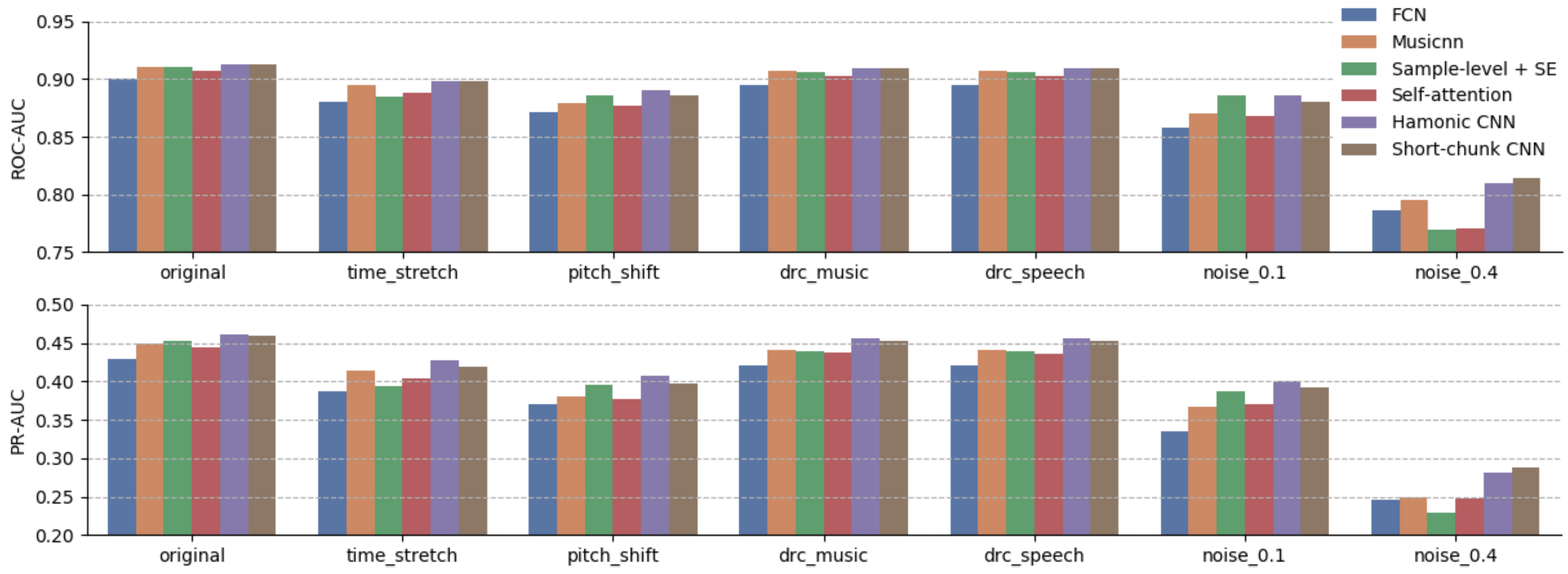} 
    \caption{Evaluations metrics with perturbed audio inputs. Dynamic range compression is shortened as ``drc" in the plot.}
    \label{fig:robustness}
\end{figure*}

Figure \ref{fig:robustness} shows performances of each model under various input deformations. Here we tested FCN, Musicnn, sample-level + SE, self-attention, Harmonic CNN, and short-chunk CNN. We followed the original input preprocessing of each model because a larger of Mel bands did not show significant effects in Section~\ref{sec:results}. CRNN is not included due to its relatively low performance. Dynamic range compression was the least influential and the white noise addition (0.4) was the most critical among the four different perturbations considered. Musicnn is robust against time stretching but it is relatively vulnerable against pitch shift. We suspect the max pooling layer over frequency axis hinders the Musicnn from learning generalized representations. Harmonic CNN and short-chunk CNN were the two best models with original data. However, Harmonic CNN showed better generalization abilities against input deformations except for the white noise addition (0.4). Sample-level CNN with SE blocks showed good performance with a small amount of noise (0.1), but it could not generalize when this amount was increased (0.4).
\section{Conclusion}\label{sec:conclusion}
In this paper, we revisit state-of-the-art automatic music tagging models and report their performances with a consistent experimental setup. 
In general, short-chunk-based approaches showed better results than models trained with larger input segments (FCN, CRNN). The design choices followed by Musicnn could show good performance on a small dataset, but it restricted the model from learning more information on larger datasets.
Sequential models (CRNN, self-attention) showed competitive results but could not outperform other models since most of tags in the datasets do not require long sequences for their identification. Interestingly, the best performing model is a simple CNN with $3 \times 3$ filters trained on short audio excerpts (short-chunk CNN). Although the original design choice of the CNN is from computer vision, it outperformed other methods except for Harmonic CNN.

We further assessed generalization abilities of models by testing perturbed inputs. We could observe a different ranking of the models in terms of their performance on each deformation. This implies that the ROC-AUC and PR-AUC scores are not enough to evaluate music tagging models. In our experiment, Harmonic CNN and short-chunk CNN consistently report better scores than other models. Specifically, Harmonic CNN showed the best generalization abilities against every deformation types except for a heavy white noise addition. 

We expect the reported results to be a useful reference for further research in automatic music tagging.
More detailed exploration of the deformations for robustness tests should be done in the future work. Also, the efficacy of data augmentation in music tagging has to be further investigated.


\begin{acknowledgments}
This work was funded by the predoctoral grant MDM-2015-0502-17-2 from the Spanish Ministry of Economy and Competitiveness linked to the Maria de Maeztu Units of Excellence Programme (MDM-2015-0502). Also we acknowledge this research has been supported by Kakao Corp.
\end{acknowledgments}

\bibliography{smc2020bib}

\begin{thebibliography}{10}
\providecommand{\url}[1]{#1}
\csname url@samestyle\endcsname
\providecommand{\newblock}{\relax}
\providecommand{\bibinfo}[2]{#2}
\providecommand{\BIBentrySTDinterwordspacing}{\spaceskip=0pt\relax}
\providecommand{\BIBentryALTinterwordstretchfactor}{4}
\providecommand{\BIBentryALTinterwordspacing}{\spaceskip=\fontdimen2\font plus
\BIBentryALTinterwordstretchfactor\fontdimen3\font minus
  \fontdimen4\font\relax}
\providecommand{\BIBforeignlanguage}[2]{{%
\expandafter\ifx\csname l@#1\endcsname\relax
\typeout{** WARNING: IEEEtran.bst: No hyphenation pattern has been}%
\typeout{** loaded for the language `#1'. Using the pattern for}%
\typeout{** the default language instead.}%
\else
\language=\csname l@#1\endcsname
\fi
#2}}
\providecommand{\BIBdecl}{\relax}
\BIBdecl

\bibitem{choi2016automatic}
K.~Choi, G.~Fazekas, and M.~Sandler, ``Automatic tagging using deep
  convolutional neural networks,'' \emph{In Proc. of the 17th International
  Society for Music Information Retrieval Conference (ISMIR)}, 2016.

\bibitem{pons2018end}
J.~Pons, O.~Nieto, M.~Prockup, E.~Schmidt, A.~Ehmann, and X.~Serra,
  ``End-to-end learning for music audio tagging at scale,'' \emph{In Proc. of
  the 19th International Society for Music Information Retrieval Conference
  (ISMIR)}, 2018.

\bibitem{lee2017sample}
J.~Lee, J.~Park, K.~L. Kim, and J.~Nam, ``Sample-level deep convolutional
  neural networks for music auto-tagging using raw waveforms,'' \emph{In Proc.
  of the 14th Sound and music computing (SMC)}, 2017.

\bibitem{kim2018sample}
T.~Kim, J.~Lee, and J.~Nam, ``Sample-level cnn architectures for music
  auto-tagging using raw waveforms,'' in \emph{Proc. of International
  Conference on Acoustics, Speech and Signal Processing (ICASSP)}.\hskip 1em
  plus 0.5em minus 0.4em\relax IEEE, 2018, pp. 366--370.

\bibitem{hu2018squeeze}
J.~Hu, L.~Shen, and G.~Sun, ``Squeeze-and-excitation networks,'' in \emph{Proc.
  of the IEEE conference on computer vision and pattern recognition (CVPR)},
  2018, pp. 7132--7141.

\bibitem{choi2017convolutional}
K.~Choi, G.~Fazekas, M.~Sandler, and K.~Cho, ``Convolutional recurrent neural
  networks for music classification,'' in \emph{Proc. of International
  Conference on Acoustics, Speech and Signal Processing (ICASSP)}.\hskip 1em
  plus 0.5em minus 0.4em\relax IEEE, 2017, pp. 2392--2396.

\bibitem{won2019toward}
M.~Won, S.~Chun, and X.~Serra, ``Toward interpretable music tagging with
  self-attention,'' \emph{arXiv preprint arXiv:1906.04972}, 2019.

\bibitem{devlin2018bert}
J.~Devlin, M.-W. Chang, K.~Lee, and K.~Toutanova, ``Bert: Pre-training of deep
  bidirectional transformers for language understanding,'' \emph{Proc. of the
  2019 Conference of the North {A}merican Chapter of the Association for
  Computational Linguistics: Human Language Technologies, Volume 1 (Long and
  Short Papers)}, 2018.

\bibitem{won2020data}
M.~Won, S.~Chun, , O.~Nieto, and X.~Serra, ``Data-driven harmonic filters for
  audio representation learning,'' \emph{In Proc. of International Conference
  on Acoustics, Speech and Signal Processing (ICASSP)}, 2020.

\bibitem{mcfee2015software}
B.~McFee, E.~J. Humphrey, and J.~P. Bello, ``A software framework for musical
  data augmentation.'' in \emph{Proc. of the 16th International Society for
  Music Information Retrieval Conference (ISMIR)}, vol. 2015, 2015, pp.
  248--254.

\bibitem{dietterich1997solving}
T.~G. Dietterich, R.~H. Lathrop, and T.~Lozano-P{\'e}rez, ``Solving the
  multiple instance problem with axis-parallel rectangles,'' \emph{Artificial
  intelligence}, vol.~89, no. 1-2, pp. 31--71, 1997.

\bibitem{mcfee2018adaptive}
B.~McFee, J.~Salamon, and J.~P. Bello, ``Adaptive pooling operators for weakly
  labeled sound event detection,'' \emph{IEEE/ACM Transactions on Audio,
  Speech, and Language Processing}, vol.~26, no.~11, pp. 2180--2193, 2018.

\bibitem{davis2006relationship}
J.~Davis and M.~Goadrich, ``The relationship between precision-recall and roc
  curves,'' in \emph{Proc. of the 23rd international conference on Machine
  learning}, 2006, pp. 233--240.

\bibitem{law2009evaluation}
E.~Law, K.~West, M.~I. Mandel, M.~Bay, and J.~S. Downie, ``Evaluation of
  algorithms using games: The case of music tagging.'' in \emph{In Proc. of
  International Society for Music Information Retrieval Conference (ISMIR)},
  2009, pp. 387--392.

\bibitem{bertin2011million}
T.~Bertin-Mahieux, D.~P. Ellis, B.~Whitman, and P.~Lamere, ``The million song
  dataset,'' \emph{In Proc. of the 12th International Society for Music
  Information Retrieval Conference (ISMIR)}, 2011.

\bibitem{choi2017effects}
K.~Choi, G.~Fazekas, K.~Cho, and M.~Sandler, ``The effects of noisy labels on
  deep convolutional neural networks for music classification,'' \emph{IEEE
  Transactions on Emerging Topics in Computational Intelligence}, 2018.

\bibitem{bogdanov2019mtg}
D.~Bogdanov, M.~Won, P.~Tovstogan, A.~Porter, and X.~Serra, ``The mtg-jamendo
  dataset for automatic music tagging,'' \emph{Machine Learning for Music
  Discovery Workshop, International Conference on Machine Learning (ICML)},
  2019.

\bibitem{bogdanov2013essentia}
D.~Bogdanov, N.~Wack, E.~G{\'o}mez~Guti{\'e}rrez, S.~Gulati, H.~Boyer,
  O.~Mayor, G.~Roma~Trepat, J.~Salamon, J.~R. Zapata~Gonz{\'a}lez, X.~Serra
  \emph{et~al.}, ``Essentia: An audio analysis library for music information
  retrieval,'' in \emph{Proc. of 14th Conference of the International Society
  for Music Information Retrieval (ISMIR)}, 2013.

\bibitem{mcfee2015librosa}
B.~McFee, C.~Raffel, D.~Liang, D.~P. Ellis, M.~McVicar, E.~Battenberg, and
  O.~Nieto, ``librosa: Audio and music signal analysis in python,'' in
  \emph{Proc. of the 14th python in science conference}, vol.~8, 2015.

\bibitem{kingma2014adam}
D.~P. Kingma and J.~Ba, ``Adam: A method for stochastic optimization,'' in
  \emph{Proc. of the International Conference on Learning Representations
  (ICLR)}, 2015.

\bibitem{bittner2017deep}
R.~M. Bittner, B.~McFee, J.~Salamon, P.~Li, and J.~P. Bello, ``Deep salience
  representations for f0 estimation in polyphonic music.'' in \emph{ISMIR},
  2017, pp. 63--70.

\bibitem{he2016deep}
K.~He, X.~Zhang, S.~Ren, and J.~Sun, ``Deep residual learning for image
  recognition,'' in \emph{Proc. of the IEEE conference on computer vision and
  pattern recognition}, 2016, pp. 770--778.

\bibitem{dolby2002standards}
E.~Dolby, ``Standards and practices for authoring dolby digital and dolby e
  bitstreams,'' \emph{Dolby Labortories, Inc}, 2002.

\end{thebibliography}

\end{document}